\documentclass[12pt]{article}

\usepackage[utf8]{inputenc}
\usepackage{cancel}
\usepackage{amssymb}
\usepackage{slashed}
\usepackage{soul}
\usepackage{tabularx}
\usepackage{xcolor}
\usepackage{booktabs}
\usepackage{subcaption} % Added by NN
\usepackage{colortbl}
\usepackage{authblk}
\DeclareUnicodeCharacter{00A0}{ }

\newcolumntype{C}{>{\centering\arraybackslash}X}
\usepackage[T1]{fontenc}
\usepackage{epsfig}
\usepackage{latexsym}
\usepackage{graphicx}
\usepackage{amsmath}
\usepackage{amsfonts}   % if you want the fonts
\usepackage{amssymb}    % if you want extra symbols
\usepackage{float}
\usepackage{bm}
\usepackage{url}
\usepackage{hyperref} 
\usepackage[nodisplayskipstretch]{setspace}
\def\order#1{\ensuremath{{\cal O}(#1)}}
\def\lsim{\raise0.3ex\hbox{$\;<$\kern-0.75em\raise-1.1ex\hbox{$\sim\;$}}}
\def\gsim{\raise0.3ex\hbox{$\;>$\kern-0.75em\raise-1.1ex\hbox{$\sim\;$}}}
\def    \beq            {\begin{equation}}
\def    \eeq            {\end{equation}}
\def    \bea           {\begin{eqnarray}}
\def    \eea           {\end{eqnarray}}

\def \mn{\mu\nu{\rm SSM}}

\def\g2{{\rm GeV}^2}
\def\sw2{sin^2 \theta_w}

\def\diag{\mathrm{diag}}

\def\a^tau{\alpha_{\tau}}

\def\beq{\begin{equation}}
\def\eeq{\end{equation}}
\def\beqa{\begin{eqnarray}}
\def\eeqa{\end{eqnarray}}

\newcommand{\tev}{\,\textrm{TeV}}

\newcommand{\newc}{\newcommand}
\newc\BR{BR}
\newc{\akappa}{A_{\kappa} }
\newc\deltagmtwo{\delta (g-2)_{\mu}} 
\newc\deltaamu{\Delta a_{\mu}}

\def\anti{\overline}

\def\la{\lambda}

\newc{\haa}{BR\(h_1\to a_1 a_1\)}
\newc{\abb}{BR\(a_1\to b\anti{b}\)}
\newc{\hbb}{BR\(h_1\to b\anti{b}\)}
\newc{\abund}{\Omega h^2}
\newc\bsgamma{b\rightarrow s \gamma }
\newc\bxsgamma{\overline{B}\rightarrow X_{s}\gamma}
\newc\brbsgamma{\BR(\overline{B}\rightarrow X_s\gamma)}

\newc{\Fermi}{\textit{Fermi}-}

\allowdisplaybreaks

\setul{0.5ex}{0.3ex}
\definecolor{Blue}{rgb}{0,0.0,1}
\setulcolor{Blue}

\usepackage{array, makecell}
\usepackage{boldline}
\usepackage[nosort]{cite}
\title{{\bf Phenomenological implications of sterile neutrinos in the $\mu\nu$SSM and dark matter}}
\author[a]{Paulina~Knees \thanks{pknees@df.uba.ar}}
\author[a,b]{Daniel~E.~López-Fogliani \thanks{daniel.lopez@df.uba.ar}}
\author[c,d]{Carlos~Mu\~noz\thanks{c.munoz@uam.es}}

\affil[a]{Instituto de Física de Buenos Aires UBA \& CONICET, Departamento de Física,
Facultad de Ciencia Exactas y Naturales, Universidad de Buenos Aires, 
1428 Buenos Aires, Argentina}
\affil[b]{
{Pontificia Universidad Católica Argentina, 
Av. Alicia Moreau de Justo 1500, 
1107~Buenos~Aires, Argentina}}
\affil[c]{Departamento de F\'{\i}sica Te\'{o}rica, Universidad Aut\'{o}noma de Madrid (UAM),
Campus~de~Cantoblanco, 28049 Madrid, Spain}
\affil[d]{Instituto de F\'{\i}sica Te\'{o}rica (IFT) UAM-CSIC, Campus de Cantoblanco, 28049 Madrid, Spain}

\date{\today}
\begin{document}
\maketitle
%%%%%%%%%%%%%%%%%%%%%%%%%%
%%%     Abstract    %%%%%%
%%%%%%%%%%%%%%%%%%%%%%%%%%
\begin{abstract}
We analyze the role of sterile neutrinos in the framework of the $\mu\nu$SSM, where the presence of right-handed neutrinos provides a simultaneous solution to $\mu$- and $\nu$-problems in supersymmetry.
We adopt a minimalistic approach, reproducing light neutrino masses and mixing angles at tree level using just two right-handed neutrinos as part of the seesaw mechanism.
A third right-handed neutrino does not contribute significantly to the mass of the three active ones, behaving as a sterile neutrino with a mass in the range keV$-$MeV. Furthermore, a sterile neutrino of about $10$~keV can be a good candidate for dark matter with a lifetime larger than the age of the Universe.
In particular,
the three-body decay to active neutrinos
gives the dominant contribution to its lifetime.
The one-loop decay to gamma and active neutrino is subdominant, but relevant for observations such as
astrophysical X-rays.
We find regions of the parameter space of the $\mu\nu$SSM, with different values of the sterile neutrino mass, fulfilling not only these constraints but also collider constraints from the Higgs sector. 
%Interestingly, the claimed $3.5$~keV line detection can also potentially be explained in the $\mu\nu$SSM with a sterile neutrino of $7$~keV mass.    

\end{abstract}
%%%%%%%%%%%%%%%%%%%%%%%%%%

%%%%%%%%%%%%%%%%%%%%%%%%%%
%%%      Keywords    %%%%%
%%%%%%%%%%%%%%%%%%%%%%%%%%


%%%%%%%%%%%%%%%%%%%%%%%%%%%%%%%%%%%%%%%%%%%%%%%%%%%%%%%%%%%%%%%%%%%%%%%%%%%%
%%%%%%%%%%%%%%%%%%   Table of Contents          %%%%%%%%%%%%%%%%%%%%%%%%%%%%
%%%%%%%%%%%%%%%%%%%%%%%%%%%%%%%%%%%%%%%%%%%%%%%%%%%%%%%%%%%%%%%%%%%%%%%%%%%%
\newpage 

\tableofcontents 
%\listoffigures
%\listoftables
%%%%%%%%%%%%%%%%%%%%%%%%%%%%%%%%%%%%%%%%%%%%%%%%%%%%%%%%%%%%%%%%%%%%%%%%%%%%

%%%%%%%%%%%%%%%%%%%%%%%%%%%%%%%%%%%%%%%%%%%%%%%%%%%%%%%%%%%%%%%%%%%%%%%%%%%%
%%%%%%%%%%%%%%%%%%         Sections                   %%%%%%%%%%%%%%%%%%%%%%
%%%%%%%%%%%%%%%%%%%%%%%%%%%%%%%%%%%%%%%%%%%%%%%%%%%%%%%%%%%%%%%%%%%%%%%%%%%%

%%%%%%%%%%%%%%%%%%%%%%%%%%%%%%%%%%%%%%%%%%%%%%%%
%%%%%%%          Introduction                %%%
%%%%%%%%%%%%%%%%%%%%%%%%%%%%%%%%%%%%%%%%%%%%%%%%
\section{Introduction}

Supersymmetry (SUSY) is one of the most interesting theories for physics beyond the standard model (SM), providing also an elegant solution to the gauge hierarchy problem. The minimal SUSY extension of the SM is the so-called Minimal Supersymmetric Standard Model (MSSM)~\cite{Nilles:1983ge,Barbieri:1987xf,Haber:1984rc,Gunion:1984yn, Martin:1997ns}.
However, it is fair to say that the MSSM has several crucial problems. Neutrinos are massless in this model implying that the MSSM itself is unable to solve the 
$\nu$-problem (the generation of neutrino masses according to experimental results~\cite{Capozzi:2017ipn,deSalas:2017kay,deSalas:2018bym,Esteban:2018azc,Abe:2019vii}).
The MSSM suffers also a naturalness problem, the so-called 
$\mu$-problem~\cite{Kim:1983dt,Bae:2019dgg}. This arises from the requirement of a mass term for the Higgs superfields in the 
superpotential, $\mu \hat H_u \hat H_d$, which must be of the order of the electroweak (EW) scale to successfully lead to EW symmetry breaking, as well as to generate Higgsino masses compatible with current experimental lower limits on SUSY particles~\cite{ParticleDataGroup:2022pth}.
However, the natural scale for $\mu$, being a SUSY mass term, is the high-energy scale expected in the theory which can be the grand unified theory (GUT) scale $M_{\text{GUT}}$, the string scale $M_{\text{string}}$, {or,
in the absence of a GUT, the Planck scale $M_{\text{Planck}}$, since gravity is always there}. In the MSSM there is no attempt to solve this problem, the $\mu$ term is just assumed to be present.

The `$\mu$~from~$\nu$' Supersymmetric Standard Model
($\mn$)~\cite{LopezFogliani:2005yw,Escudero:2008jg} is a highly predictive alternative model to the MSSM.
The $\mn$ solves simultaneously the
$\mu$-problem 
and the $\nu$-problem without
the need to introduce additional energy scales beyond the SUSY-breaking scale (See Ref.~\cite{Lopez-Fogliani:2020gzo} for a recent review of the $\mn$ and Ref.~\cite{Biekotter:2021rak} for {the latest} 
%a
vacuum structure analysis).
Summarizing, 
in the $\mn$ the particle content of the MSSM
is extended by right-handed neutrino superfields $\hat \nu^c_i$, allowing gauge invariant couplings with the Higgses of the type
$\lambda_{i} \, \hat \nu^c_i\,\hat H_u \hat H_d$ in the superpotential.
They generate dynamically the $\mu$-term after the   symmetry breaking induced by the soft SUSY-breaking terms which are in the ballpark of a TeV, since then
the right sneutrinos 
$\widetilde \nu_{iR}$ develop vacuum expectation values (VEVs) also of the order of TeV. Thus, $\mu=\la_i \langle \widetilde \nu_{iR}\rangle$.
In addition,
Dirac Yukawa couplings between left-handed (LH) neutrinos $\nu_{iL}$ and right-handed (RH) neutrinos $\nu_{iR}$, $Y^{\nu}_{ij} \, \hat H_u\, \hat L_i \, \hat \nu^c_j$, as well as
couplings among RH neutrinos themselves,
$\kappa{_{ijk}} \hat \nu^c_i\hat \nu^c_j\hat \nu^c_k$, are allowed, with
the latter
generating effective EW-scale Majorana masses for RH neutrinos,
${\mathcal M}_{ij}\sim\kappa_{ijk}\langle \widetilde \nu_{kR}\rangle$. 
Both types of couplings
are therefore 
instrumental 
in solving the $\nu$-problem through an EW-scale seesaw. They
can accommodate at tree level the correct neutrino masses and mixing angles with
$Y^{\nu}_{ij} \lsim 10^{-6}$~\cite{LopezFogliani:2005yw,Escudero:2008jg,Ghosh:2008yh,Bartl:2009an,Fidalgo:2009dm,Ghosh:2010zi,Liebler:2011tp}. 
Actually, this is even possible with diagonal Yukawa couplings, i.e.\
$Y^{\nu}_{ij}=Y^{\nu}_{i}\delta_{ij}$~\cite{Ghosh:2008yh,Fidalgo:2009dm}.
Having 
a {EW seesaw} also avoids the introduction of {\it ad-hoc} high-energy scales in the model, as it occurs e.g. in the case of a GUT seesaw.  
Thus, the only scale in the $\mn$ is the EW symmetry breaking scale.
Last but not least, baryon asymmetry of the Universe might be realized in the $\mn$ through EW baryogenesis~\cite{Chung:2010cd}.

On the other hand, the search for SUSY at accelerators has been focused mainly on prompt signals with missing transverse energy (MET) from neutralinos inspired in models such as the
MSSM or the Next-to-MSSM (NMSSM)~\cite{Maniatis:2009re,Ellwanger:2009dp}.
In these models, 
the $Z_2$ discrete symmetry $R$-parity (+1 for particles and -1 for SUSY particles) is imposed by hand in order to avoid fast proton decay.
This $R$-parity conservation (RPC) has the consequence that SUSY particles must appear in pairs. Thus, 
the lightest supersymmetric particle (LSP) is stable and a potential candidate for dark matter (DM). Since it must be a neutral particle, the only candidate in the MSSM/NMSSM is the neutralino (a mixture of bino, neutral wino and Higgsinos).
In this RPC framework of searching for prompt signals with MET from neutralinos at accelerators, significant lower bounds on SUSY particle masses have been obtained~\cite{ParticleDataGroup:2022pth}. In particular, strongly interacting  SUSY particles have to have masses above about 1 TeV, whereas the bound for the weakly interacting SUSY particles is about 100 GeV (with the exception of the bino-like neutralino which is basically not constrained due to its small pair production cross section).

In contrast to the MSSM/NMSSM, $R$-parity (and lepton number) is not conserved in the $\mn$.
The simultaneous presence in the superpotential of the three terms discussed above makes it impossible to assign $R$-parity charges consistently to 
%the RH neutrinos,
$\nu_{iR}$, 
thus producing explicit $R$-parity violation (RPV), although harmless for proton decay. 
Note that in the limit of neutrino Yukawa couplings
$Y^{\nu}_{ij} 
\to 0$, $\hat \nu^c_i$ can be identified in the superpotential 
as 
pure singlet superfields without lepton number, similar to the singlet of the NMSSM, and therefore 
$R$-parity is restored.
Thus, $Y^{\nu}$ are the parameters which control the amount of RPV in the $\mn$, and as a consequence
this violation is small
since the size of $Y^{\nu}_{ij}\lsim 10^{-6}$ is determined by the EW-scale 
seesaw of the $\mn$, as discussed above.
This novel RPV framework leads to a completely different
phenomenology, since SUSY particles do not appear in pairs and therefore the LSP is not stable, decaying to SM particles.
Now, depending on the LSP mass, signals at accelerators are
characterized not only by prompt decays of the LSP, but also by displaced
decays of the order of mm$-$m due to
its low decay width produced by the smallness of neutrino masses.
These decays produce distinct signals with multi-leptons/jets/photons with small/moderate MET from 
neutrinos~\cite{Ghosh:2017yeh,Lara:2018rwv,Lara:2018zvf,Kpatcha:2019gmq,Kpatcha:2019pve,Heinemeyer:2021opc,Kpatcha:2021nap}.
Besides, since the LSP is not stable all SUSY particles, not only the neutralino, can be candidates for the LSP: squarks, gluinos, charginos, charged sleptons, sneutrinos.
These features imply that the extrapolation of the usual MSSM/NMSSM bounds on SUSY particle masses~\cite{ParticleDataGroup:2022pth} to the $\mn$ is not applicable~\cite{Ghosh:2017yeh,Lara:2018rwv,Lara:2018zvf,Kpatcha:2019gmq,Kpatcha:2019pve,Heinemeyer:2021opc,Kpatcha:2021nap}.

A similar conclusion concerning the phenomenology at accelerators is obtained when comparing the $\mn$ with other RPV models proposed in the literature~\cite{Barbier:2004ez}, such as the bilinear RPV model (BRPV) where the bilinear terms $\mu_i \hat H_u\, \hat L_i $ are added to the superpotential to generate neutrino masses,
and the trilinear RPV models (TRPV) where 
the lepton-number violating couplings
$\lambda_{ijk} \hat L_i \hat L_j \hat e^c_k + \lambda'_{ijk}\hat L_i \hat Q_{j} \hat d^{c}_{k}$ or the baryon-number violating couplings 
$\lambda''_{ijk}\hat d^c_{i} \hat d^c_{j} \hat u^c_{k}$, are added.
Besides, in these models neutrino masses can only arise through loop corrections, and they do not attempt to solve the $\mu$-problem.

The low decay width of the LSP due to the smallness of neutrino masses,
is also related to the existence of candidates for (decaying) DM in the $\mn$.
This is the case of 
the gravitino~\cite{Choi:2009ng,GomezVargas:2011ph,Albert:2014hwa,GomezVargas:2017,Gomez-Vargas:2019mqk}, or the axino~\cite{Gomez-Vargas:2019vci}, with lifetimes longer than the age of the Universe, since
their interactions are not only supressed by the Planck or the Peccei-Quinn scale, but also by the small RPV parameters.
Contrary to the MSSM/NMSSM where the stable neutralino can be detected in accelerator experiments as well as DM direct detection experiments~\cite{Munoz:2003gx},
or an extension of the NMSSM where the right sneutrino can be the DM~\cite{Cerdeno:2008ep,Lopez-Fogliani:2021qpq},
in the $\mn$ these types of detection techniques are not possible. The decaying DM candidates, gravitino and axino, are so weakly interacting that they can only be observed in indirect detection experiments such as MeV-GeV gamma-ray telescopes, through their decays into photon and neutrino~\cite{Choi:2009ng,GomezVargas:2011ph,Albert:2014hwa,GomezVargas:2017,Gomez-Vargas:2019mqk,Gomez-Vargas:2019vci}.
Thus displaced decay signals at accelerators and/or gamma-ray signals at indirect detection experiments are clues for the $\mn$ compared to other models.

On the other hand, a sterile neutrino is an intriguing particle that could be detectable through its mixing with the active neutrinos. In particular,
with appropriate mass and couplings, it is a well motivated DM candidate with a lifetime longer than the age of the Universe.
The viability of the sterile neutrino as decaying DM, and also its possible detection through different signals, has been studied extensively in 
recent years (for a review, see e.g. Ref.~\cite{Drewes:2016upu} and references therein).
Since
the $\mn$ includes naturally RH neutrinos, we might use some of them to genererate light neutrino masses, whereas others can be used as sterile neutrinos. 
Thus in the {natural} case of three RH neutrinos that we will adopt here, one can use one of them with a mass in the range of keV$-$MeV, without a significant participation in the seesaw mechanism, as a sterile neutrino.
This is the aim of this work, to carry out the first analysis of the phenomenology of the sterile neutrino in the $\mn$.
Unlike the other DM candidates, gravitino and axino, that can be detected through gamma-ray signals as discussed above, the sterile neutrino can be observed through X-rays.

The paper is organized as follows. 
In Section~\ref{sec:munu}, we will briefly review the $\mn$ and its relevant parameters for the analysis of the neutrino sector, and in particular we will apply this information to learn how to use one of the RH neutrinos of the model as a sterile neutrino. In Section \ref{sec:decay}, we will study the relevant decay modes of the sterile neutrino to determine its lifetime, and therefore its viability as a DM candidate with a mass in the range of keV.
Section \ref{sec:xray} is dedicated to discuss the viability of its detection via radiative decay modes, as well as the associated exclusion limits. 
In Section \ref{sec:results}, we will carry out a numerical analysis of the parameter space 
%\st{to obtain benchmark points} 
{to obtain a representative set of parameters} compatible with a sterile neutrino reproducing at the same time data on neutrino masses and mixing angles.
Finally, our conclusions are left for Section~\ref{sec:conclusions}.

%%%%%%%%%%%%%%%%%%%%%%%%%%%%%%%%%%%%%%%%%%%%%%%%
%%%%%           Sterile                      %%%
%%%%%%%%%%%%%%%%%%%%%%%%%%%%%%%%%%%%%%%%%%%%%%%%
\section{Sterile neutrinos in the $\mn$}
\label{sec:munu}

\subsection{The neutrino sector in the $\mn$} 

%In the $\mn$~\cite{LopezFogliani:2005yw,Escudero:2008jg}, the particle content of the MSSM
%is extended by RH neutrino superfields $\hat \nu^c_i$. 
The simplest superpotential of the $\mn$ with three RH neutrinos
is~\cite{LopezFogliani:2005yw,Escudero:2008jg,Ghosh:2017yeh}
\bea
W &=&
\epsilon_{ab} \left(
Y^e_{ij}
%Y^e_{ij} 
\, \hat H_d^a\, \hat L^b_i \, \hat e_j^c +
Y^d_{ij} 
%Y^d_{ij} 
\, 
%\delta_{\alpha\beta}\, 
\hat H_d^a\, \hat Q^{b}_{i} \, \hat d_{j}^{c} 
+
Y^u_{ij} 
%Y^u_{ij} 
\, 
%\delta_{\alpha\beta}\, 
\hat H_u^b\, \hat Q^{a}_i
%_{i\alpha} 
\, \hat u_{j}^{c}
\right)
\nonumber\\
% &+&
% \epsilon_{ab} Y^{\nu}_{ij} \, \hat H_u^b\, \hat L^a_i \, \hat \nu^c_j -
&+& 
\epsilon_{ab} \left(
Y^{\nu}_{ij} 
%Y^{\nu}_{i} 
\, \hat H_u^b\, \hat L^a_i \, \hat \nu^c_j
-
%\epsilon_{ab}
\lambda_i \, \hat \nu^c_i\, \hat H_u^b \hat H_d^a
\right)
+
\frac{1}{3}
\kappa_{ijk}
\hat \nu^c_i\hat \nu^c_j\hat \nu^c_k,
\label{superpotential}
\eea
where the summation convention is implied on repeated indices, with $i,j,k=1,2,3$ the usual family indices of the SM
and $a,b=1,2$ $SU(2)_L$ indices with $\epsilon_{ab}$ the totally antisymmetric tensor, $\epsilon_{12}= 1$. 
%and $i,j,k=1,2,3$ the usual family indices of the SM.
The first three terms are the usual Dirac Yukawa couplings for quarks and charged leptons.
As discussed in the Introduction,
the last three terms 
are characteristic of the $\mn$. 
It is worth noting here that because of the VEVs acquired by the RH sneutrinos after the EW symmetry breaking, the neutrino Yukawa couplings also generate dynamically the bilinear terms $\mu_i \hat H_u\, \hat L_i $
used in the BRPV,
with $\mu_i=Y^{\nu}_{ij}\langle \widetilde \nu_{jR}\rangle$.

With the choice of CP conservation,\footnote{The $\mn$ with spontaneous CP violation was studied in Ref.~\cite{Fidalgo:2009dm}.} the neutral components of the Higgs doublet fields $H_d$ and $H_u$, and the left and right sneutrinos $\widetilde\nu_{iL}$ and $\widetilde\nu_{iR}$
develop real VEVs denoted by:  
\begin{eqnarray}
\langle H_{d}^0\rangle = \frac{v_{d}}{\sqrt 2},\quad 
\langle H_{u}^0\rangle = \frac{v_{u}}{\sqrt 2},\quad 
\langle \widetilde \nu_{iR}\rangle = \frac{v_{iR}}{\sqrt 2},\quad 
\langle \widetilde \nu_{iL}\rangle = \frac{v_{iL}}{\sqrt 2}.
\end{eqnarray}
 Then, the higgsino mass parameter $\mu$, 
 the bilinear couplings $\mu_i$ 
 and Dirac and Majorana masses are given by:
\bea
\mu=\la_i \frac{v_{iR}}{\sqrt 2}, \;\;\;\;
\mu_i=Y^{\nu}_{ij}  \frac{v_{jR}}{\sqrt 2}, \;\;\;\;
m_{{\mathcal{D}_{ij}}}= Y^{\nu}_{ij} 
\frac{v_u}{\sqrt 2}, \;\;\;\;
{\mathcal M}_{ij}
={2}\kappa_{ijk} \frac{v_{kR}}{\sqrt 2}.
\label{mu2}    
\eea

%On the other hand, the simultaneous presence of the three terms discussed above makes it impossible to assign $R$-parity charges consistently to $\nu_{iR}$, 
%thus producing explicit $R$-parity violation (RPV), although harmless for proton decay. 
%Note that in the limit of neutrino Yukawa couplings
%$Y^{\nu}_{ij} 
%\to 0$, $\hat \nu^c_i$ can be identified in the superpotential 
%as 
%pure singlet superfields without lepton number, similar to the singlet of the NMSSM, and therefore 
%$R$-parity is restored.
%Thus, $Y^{\nu}$ are the parameters which control the amount of RPV in the $\mn$, and as a consequence
%this violation is small
%since the size of $Y^{\nu}_{ij}\lsim 10^{-6}$ is determined by the EW-scale 
%seesaw of the $\mn$~\cite{LopezFogliani:2005yw,Escudero:2008jg}.

As discussed in the Introduction, 
$v_{iR}\sim {\order{1 \tev}}$, 
because of the right sneutrino minimization equations which are
determined by the soft SUSY-breaking terms~\cite{LopezFogliani:2005yw,Escudero:2008jg,Ghosh:2017yeh}.
However, the small values of $Y^{\nu}_{ij}\lsim 10^{-6}$ determined by the EW-scale seewsaw of the $\mn$, 
drive $v_{iL}$ to small values because of the proportional contributions to
$Y^{\nu}_{ij}$ appearing in their minimization equations. {A rough} estimation gives
$v_{iL}\lsim m_{{\mathcal{D}_{jk}}}\lsim 10^{-4}$ GeV.
% the contrary, tipically $v_{iR}\sim {\order{1 \tev}}$, 
%because of the right sneutrino minimization equations which are
%determined by the soft SUSY-breaking terms~\cite{LopezFogliani:2005yw,Escudero:2008jg,Ghosh:2017yeh}.

As a consequence of the RPV of the $\mn$, 
the MSSM neutralinos mix with the LH and RH neutrinos, thus 
the neutral fermions have the flavor composition 
${\psi^0}^T=({(\nu_{iL})^c}^*,\widetilde B^0,\widetilde W^0,\widetilde H_{d}^0,\widetilde H_{u}^0,\nu_{iR}^*)$.
The mass terms in the Lagragian are given by $-\frac{1}{2} {\psi^0}^T m_{{\psi^0}} {\psi^0}$ + h.c., 
where $m_{{\psi^0}}$ is a $10\times 10$ (symmetric) neutrino/neutralino mass matrix   
with the structure of a generalized EW-scale seesaw given by~\cite{Escudero:2008jg,Ghosh:2017yeh}:
\begin{equation}m_{\psi^0}=\begin{pmatrix}
 0_{3\times 3} & m^T \\
 m & {M}_{7\times7}
    \end{pmatrix}.
\label{eq:mass matrix}
\end{equation}    
%In equation \ref{eq:mass matrix}, 
Here $m$ is a $3\times7$ submatrix containing the mixing of LH neutrinos with MSSM neutralinos
and RH neutrinos,
\begin{equation}m^T=\begin{pmatrix}
-\frac{1}{{2}}g'{v}_{iL}  & -\frac{1}{{2}} g{v}_{iL}  & 0_{3\times 1} & \mu_i & 
m_{{\mathcal{D}_{ij}}}
     \end{pmatrix},
\label{eq:mchica}     
\end{equation}
where the EW gauge couplings are estimated at the $m_Z$ scale by
$e=g\sin\theta_W=g'\cos\theta_W$.
${M}$ is a $7\times 7$ submatrix containing the mixing of MSSM neutralinos with RH neutrinos, in addition to the mixing between MSSM neutralinos themselves and RH neutrinos themselves, 
\begin{equation}{M}=
%\resizebox{.9\textwidth}{!}
\begin{pmatrix}
M_1 & 0 & -\frac{1}{2} g'v_d & \frac{1}{2}g' v_u  & 0_{1\times 3} \\
0 & M_2 & \frac{1}{2}g v_d & -\frac{1}{2} g v_u  & 0_{1\times 3}  \\
-\frac{1}{2}g' v_d & \frac{1}{2}g v_d & 0 & -\mu  & -\lambda_j v_u \frac{1}{\sqrt 2}  \\
\frac{1}{2}g' v_u & -\frac{1}{2}g v_u & -\mu & 0 & \left(-\lambda_j {v_d} + Y^{\nu}_{kj} {v_{kL}}\right) \frac{1}{\sqrt 2}\\
0_{3\times 1} & 0_{3\times 1} & -\lambda_j {v_u}\frac{1}{\sqrt 2}  &
  \left(-\lambda_j {v_d} + Y^{\nu}_{kj} {v_{kL}}\right)\frac{1}{\sqrt 2} 
   & {\mathcal M}_{ij} 
    \end{pmatrix}.
\label{eq:mgrande}
\end{equation}

Regarding the seesaw mechanism, note that the entries of $M$
are expected to be of the order of the EW/SUSY scale, since they contain
the bino and wino soft masses, $M_1$ and $M_2$, the $\mu$ term, the mixing of Higgsinos with RH neutrinos which is essentially determined by $\lambda_i v_u$ and
$\lambda_i v_d$, and the self mixing of the RH neutrinos which is determined by
the Majorana masses  ${\mathcal M}_{ij}$  of Eq.~(\ref{mu2}) which are $\sim {\order{1 \tev}}$.
Here we are assuming values of the couplings $\lambda_i, \kappa_{ijk}\sim {\order{1}}$.
On the contrary, the entries of the matrix $m$ are much smaller, since the bilinear mass terms $\mu_i$ and Dirac neutrino masses $m_{{\mathcal{D}_{ij}}}$ are $\lsim 10^{-4}$ GeV as discussed
above, and the other entries are similarly small being determined by $g v_{iL}$ and $g' v_{iL}$.

Now, in a first approximation the effective 
active-neutrino mixing mass matrix 
can be written as~\cite{Fidalgo:2009dm}
\begin{equation}
    m^{\text{eff}}_\nu=-m^T\ {M}^{-1}\ m,
\end{equation}
and, after diagonalization by an unitarity transformation $U_{\text{\tiny{PMNS}}}$, one obtains
\begin{equation}
    U_{\text{\tiny{PMNS}}}^T\ m^{\text{eff}}_\nu\ U_{\text{\tiny{PMNS}}}=\diag\ (m_{\nu_1},m_{\nu_2},m_{\nu_3}).
\end{equation}
We conclude therefore that the relevant independent parameters
in the neutrino sector are:
\begin{equation}
    v_{iL},\; v_{iR},\; \tan \beta,\; Y^{\nu}_{ij},\; \kappa_{i},\; \lambda_i,\; M_1,\; M_2,
    \label{eq:independent params}
\end{equation}
where we have assumed $\kappa_{iii}\equiv \kappa_i$ and vanishing otherwise, which
is sufficient for the purposes of our numerical analysis below.
Also we have determined $v_u$ and $v_d$ using $\tan\beta\equiv v_u/v_d$ and $v_d\approx v/\sqrt{\tan^2\beta+1}$, since the SM Higgs VEV, $v^2 = v_d^2 + v_u^2 + \sum_i v^2_{iL}={4 m_Z^2}/{(g^2 + g'^2)}\approx$ (246 GeV)$^2$.

To give us a qualitative idea of the values of neutrino masses in the model, in the simplifying situation of universal $v_{iR}=v_R$, $\lambda_i=\lambda$, $\kappa_{i}=\kappa$, and diagonal $Y^{\nu}_{ij}=Y^{\nu}_{i}\delta_{ij}$, one can obtain the approximate formula~\cite{Fidalgo:2009dm}:
\begin{eqnarray}
\label{Limit no mixing Higgsinos gauginos}
(m^{\text{eff}}_{\nu})_{ij} 
\approx
\frac{m_{{\mathcal{D}_i}} m_{{\mathcal{D}_j}} }
{3{\mathcal{M}}}
                   \left(1-3 \delta_{ij}\right)
                   -\frac{v_{iL}v_{jL}}
                   {4M_g}, \;\;\;\;\;\;\;\;
        \frac{1}{M_g} \equiv \frac{g'^2}{M_1} + \frac{g^2}{M_2},    
\label{neutrinoph2}
  \end{eqnarray}     
with 
$m_{{\mathcal{D}_{i}}}= Y^{\nu}_{i} {v_u}/{\sqrt 2}$ and
${\mathcal M}
={2}\kappa {v_{R}}/{\sqrt 2}$.
{Of the three terms in Eq.~(\ref{neutrinoph2}),
the first two 
are generated through the mixing 
of $\nu_{iL}$ with 
$\nu_{iR}$-Higgsinos, and the third one 
also include the mixing with gauginos.
{These are the so-called $\nu_{R}$-Higgsino seesaw and gaugino seesaw, respectively~\cite{Fidalgo:2009dm}.}
}
As can be straightforwardly understood from this equation, for values of the couplings and parameters $\lambda, \kappa\sim 1$,
 $Y_{\nu_{ij}}\lsim 10^{-6}$, 
$v_{iL}\lsim 10^{-4}$, and $M_1, M_2, v_{iR}\sim 1 \tev$, as discussed above, light (basically LH)
neutrino masses $\lsim 0.1$ eV, accompanied by RH neutrino masses $\sim 1$ TeV, can easily be obtained.

This result also gives us an idea that the values of several of the parameters of the model must be chosen differently if we want to obtain a neutrino mass $\sim$ keV$-$MeV, as required for a sterile neutrino.
This is the discussion we will focus on below.

\subsection{Sterile neutrinos} 
\label{sterile}

Sterile neutrinos are singlets under the SM gauge group, and, as discussed in the Introduction, 
they are very interesting particles for phenomenology beyond the SM. 
Given our review above of neutrino physics in the $\mn$, we will analyze the simplest case of converting one of the Majorana RH neutrinos into a sterile neutrino with a mass $\sim$ keV$-$MeV, modifying the values of some of the input parameters of the model.
In this case, we are effectively using only two RH neutrinos in the seesaw mechanism, and therefore we will need to work with off-diagonal terms in the neutrino Yukawa matrix in order to reproduce correctly neutrino data {at tree level}.

The first $4\times 4$ entries of the 
matrix $M$ in Eq.~(\ref{eq:mgrande}) can be identified with the entries of the MSSM neutralino matrix. As already mentioned, their contributions are naturally of order EW/SUSY scale since they contain EW gaugino masses and the effective $\mu$ parameter (which is experimentally bounded by the chargino mass). Nevertheless, RH neutrino masses depend on the values chosen for the $\kappa_{ijk}$ parameters in ${\mathcal M}_{ij}$  as well as on
the values of $\lambda_i$ and $Y^{\nu}_{ij}$ controlling their mixtures with MSSM higgsinos and active neutrinos.

Thus, for the purpose of obtaining one light sterile neutrino $\nu_S$, it is necessary to decouple one of the RH neutrinos from the seesaw mechanism,
reducing the value of one of the diagonal Yukawa couplings up to $Y^{\nu}_{ii}\sim 10^{-13}$.
Now, in order to achieve the desired mass scale in the range of {1 keV$-$100 keV}, we 
obtain from the Majorana mass in Eq.~(\ref{mu2}) the relation
 \bea
m_{\nu_s} \approx \sqrt{2} \; \kappa_2 v_{2R}\approx 10^{-6}-{10^{-4}}\ \text{GeV},
\label{eq:sterile mass}
\eea
where for the sake of definiteness we have chosen the second family of RH neutrinos to be the sterile neutrino, $\nu_{2R}\equiv \nu_s$. Our results can be easily extrapolated to the case of using the first or the third family as the ones for the sterile neutrino.
From Eq.~(\ref{eq:sterile mass}), RH sneutrino VEVs $v_{iR}\sim 10^3$ GeV imply that
we have to use a value of {$\kappa_2\sim 10^{-9}-10^{-7}$}. 
Summarizing, two of the independent parameters of 
Eq.~(\ref{eq:independent params}) must be fixed to the following values:
 \begin{equation}
 Y^{\nu}_{22}\sim 10^{-13}, \quad
 \kappa_{2}\sim 10^{-9}-{10^{-7}.}
 \label{eq:values}
 \end{equation}

As mentioned above when discussing the matrix in Eq.~(\ref{eq:mgrande}),
higgsinos and RH neutrinos are mixed, 
and as a consequence $\lambda_2$ must be chosen with an appropriate value to obtain a viable mixture {between higgsinos and the sterile neutrino.} {Given that in general the order of magnitude of the mixtures with active neutrinos  is $\sim 10^{-7}$, and we need the sterile neutrino to be almost a pure state, it is reasonable to take this value for the mixing. Thus we can roughly approximate the mixing by the ratio $\lambda_2 v_u/\mu$, implying } 
 \begin{equation}
   \lambda_2 \sim 10^{-7}.
    \label{eq:independent}
\end{equation}

Using the values of the parameters of Eqs.~(\ref{eq:values}) and~(\ref{eq:independent}), it is possible to find regions of the model with a mass of the sterile state not only in the {1 keV$-$100 keV range, but also with a lifetime long enough to be a DM candidate.}
We will focus on this latter possibility in the next section, and the numerical analysis will be carried out in
Section~\ref{sec:results}.

 {The hierarchies between the values of the parameters of the model, $Y^\nu$'s, $\lambda_i$'s,
 $k_i$'s are an unavoidable consequence of imposing a light sterile neutrino. One might think that this situation is not so natural, although it is true that something similar occurs in the SM with the values of Yukawa couplings. In the context of these always uncertain naturalness arguments, it worth noting that we are working here for simplicity with three families of RH neutrino superfields, but in fact this is an arbitrary number
in the model. Thus, with more than three families one could argue that such hierarchies are not unexpected.}

Let us finally discuss the range of masses of the scalar and pseudoscalar partners of the sterile neutrino.
{As a consequence of RPV violation, 
the Higgses mix with left and right sneutrinos giving rise to eight scalar states $h^0_\alpha$,
$\alpha=1,...,8$, and seven pseudoscalar states $A^0_{\alpha'}$, $\alpha'=1,...,7$ (since the remaining pseudoscalar is the Goldsone boson).
All their masses are in the EW/SUSY range $\sim 10^2-10^3$~GeV, except for the scalar and pseudoscalar partners of the sterile neutrino, $h^0_1\equiv\widetilde{\nu}_s^{\mathcal{R}}$ and
$A^0_1\equiv\widetilde{\nu}_s^{\mathcal{I}}$, which are lighter as expected from the above values of the associated parameters.}
Using the results of Refs.~\cite{Escudero:2008jg,Ghosh:2017yeh}, from the minimization equation with respect to $v_{2R}$ their physical masses can be written
in a good approximation as:
\bea
m^2_{\widetilde{\nu}_s^{\mathcal{R}}} &\approx& T^{\lambda}_2 \frac{v_d v_u}{\sqrt{2} v_{2R}} + T^{\kappa}_2 \frac{v_{2R}}{\sqrt 2} + m^2_{\nu_s}  
%\nonumber \\
 =  A^{\lambda}_2 \lambda_2 \frac{v_d  v_u}{\sqrt{2} v_{2R}} + \frac{1}{2}A^{\kappa}_2 m_{\nu_s}+ m^2_{\nu_s} , \\
% & \sim &  \lambda_2 \; A_{\lambda_2} \frac{v_d \; v_u}{\sqrt{2} \; v_{2R}}, \\
%\label{eq:sneutrilo sterile mass R}
%\eea
%\bea
m^2_{\widetilde{\nu}_s^{\mathcal{I}}} 
%& \approx & m^2_{\tilde{\nu}_S^{\mathcal{R}}} - 2 \; \left( \sqrt{2} \; \kappa_2  \; A_{\kappa_2}    v_{2R}
%- \lambda_2 \; \kappa_2 \; v_d \; v_u \right) -  m^2_{\nu_S}  \nonumber \\
%&  \approx  & m^2_{\tilde{\nu}_S^{\mathcal{R}}} - 2 \; m_{\nu_S}  \; A_{\kappa_2} \;  
%+ \sqrt{2} \; \lambda_2 \; m_{\nu_S}  \; \frac{v_d \; v_u}{v_{2R}} \nonumber \\ 
&  \approx  & m^2_{\widetilde{\nu}_s^{\mathcal{R}}} - 2 \sqrt 2\ T^{\kappa}_2v_{2R}
-m^2_{\nu_s}\
=  A^{\lambda}_2 \lambda_2 \frac{v_d  v_u}{\sqrt{2} v_{2R}} 
%T_{\lambda_2} \frac{v_d v_u}{\sqrt{2} v_{2R}} 
%-\frac{3}{\sqrt 2}\ T_{\kappa_2}v_{2R}
-\frac{3}{2}  A^{\kappa}_2 m_{\nu_s},
\label{eq:sneutrilo sterile mass R}
\eea
where in the last equalities the supergravity breaking as the origin of the soft SUSY-breaking terms is assumed, with
the trilinear parameters proportional to their corresponding couplings, i.e. $T^{\kappa}_2= A^{\kappa}_2 \kappa_2$, $T^{\lambda}_2= A^{\lambda}_2 \lambda_2$. 

Clearly, if the terms with $A^{\kappa}_2$ dominate the sum, the scalar (pseudoscalar) would be tachyonic for $A^{\kappa}_2$ negative (positive). Thus we have to choose the parameters in such a way that the term proportional to 
$A^{\lambda}_2$ is the dominant one.
Assuming $ \lambda_2 \;, A^{\lambda}_2  >0 $,  we can make a good estimation for $m^2_{\widetilde{\nu}_s^{\mathcal{I}}}$ and $ m^2_{\widetilde{\nu}_s^{\mathcal{R}}} $, obtaining for the $A^{\kappa}_2 < 0$ case,
 \bea
m^2_{\widetilde{\nu}_s^{\mathcal{I}}} \gtrsim m^2_{\widetilde{\nu}_s^{\mathcal{R}}} \sim  \lambda_2 \; A^{\lambda}_2 \frac{v_d \; v_u}{\sqrt{2} \; v_{2R}} > \frac{1}{2} \; m_{\nu_s}  \;| A^{\kappa}_2|,
\label{eq:sneutrilo sterile mass R2}
\eea
and for the $A^{\kappa}_2 > 0$ case,
 \bea
m^2_{\widetilde{\nu}_s^{\mathcal{R}}} \gtrsim m^2_{\widetilde{\nu}_s^{\mathcal{I}}} \sim  \lambda_2 \; A^{\lambda}_2 \frac{v_d \; v_u}{\sqrt{2} \; v_{2R}} > \frac{3}{2} \; m_{\nu_s}  \; A^{\kappa}_2.
\label{eq:sneutrilo sterile mass R3}
\eea
 It is important to notice here that the scalar and the pseudoscalar partner of the sterile neutrino are light compared with the SM Higgs but much heavier than the sterile neutrino, 
 $m_{\widetilde{\nu}_s^{\mathcal{I}}} \sim m_{\widetilde{\nu}_s^{\mathcal{R}}} >> m_{\nu_s}$.
{From the above equations, using $A_2^\lambda\sim 10^3$ GeV one obtains $m_{\widetilde{\nu}_s^{\mathcal{I,R}}}\sim 10$ MeV.}
However, 
since they are dominated by the RH sneutrino composition their production at colliders is not observable.

%%%%%%%%%%%%%%
%   Decay    %
%%%%%%%%%%%%%%
\section{Sterile neutrinos as dark matter
}
\label{sec:decay}

In this section, we will study the viability of the sterile neutrino as a DM candidate in the $\mn$.
It has two relevant decay modes. One is the three-body decay to active neutrinos
$\nu_s\to \nu \nu \nu $, and the second one is the radiative decay to active neutrino and gamma $\nu_s \to \nu \gamma$. Given that the
latter is a one-loop diagram, it has a lower decay width than the former. 
Thus it is sufficient for our purposes to consider only the three-body decay process for the calculation of the lifetime.\par 

\begin{figure}[t!]
    \centering
    \includegraphics[scale=0.5]{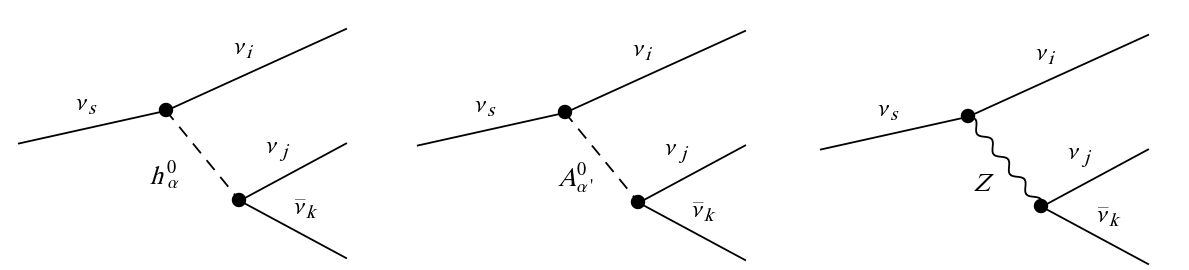}
    \caption{Three-body decays of a sterile neutrino to neutrinos mediated by scalars $h^0_\alpha$,
$\alpha=1,...,8$, 
with
$h^0_1\equiv\widetilde{\nu}_s^{\mathcal{R}}$,
pseudoscalars $A^0_{\alpha'}$, $\alpha'=1,...,7$, 
with
$A^0_1\equiv\widetilde{\nu}_s^{\mathcal{I}}$, 
and a $Z$ boson. 
    }
    \label{fig:treelevel}
\end{figure}

In particular, we have the three Feynman diagrams shown in Fig \ref{fig:treelevel} contributing to the decay width of the sterile neutrino. 
In addition to the one with the $Z$ boson as a mediator, in the $\mn$ we have also the possibility of the neutral scalars and pseudoscalars as mediators. 
{As discussed in the previous section, all these masses are in the range $\sim 10^2-10^3$~GeV, except for the scalar and pseudoscalar partners of the sterile neutrino, $\widetilde{\nu}_s^{\mathcal{R}}$ and
$\widetilde{\nu}_s^{\mathcal{I}}$ respectively, which are lighter $\sim 10^{-2}$ GeV.}
But even these contributions to the decay width are supressed with 
respect to the $Z$ one.
{Note in this respect that the latter contribution arises mainly from the sterile-active neutrino mixing in the incoming line. However, the main contribution mediated by scalar/pseudoscalar Higgses has, in addition to the sterile-active neutrino mixing in one of the outgoing lines, the suppression of the two neutrino Yukawa couplings involved in both vertices.}

For the decay width mediated by the $Z$ boson one has:
\begin{equation}
    \Gamma (\nu_s\to \nu_i \nu \bar{\nu})=\sum_{j,k}\Gamma(\nu_s \to \nu_i \nu_j \bar{\nu}_k)=\sum_{j,k} \frac{G_F^2 m_{\nu_s}^5}{6 \pi^3} \; \left(O_{si} O_{jk}\right)^2,
    \label{eq:gamma_3nu}
\end{equation}
where $G_F=\sqrt 2 g^2/8m^2_W$ is the Fermi constant, and
\begin{equation}
    O_{pq}=-\frac{1}{2}U^V_{p6}U^{V}_{q6}+\frac{1}{2}U^V_{p7}U^{V}_{q7}-\frac{1}{2}\sum_{r=1}^3 U^V_{pr}U^{V}_{qr},
\label{eq:Omixing}
\end{equation}
with $U^V$ the matrix which diagonalizes the mass matrix 
 of the 
neutral fermions in Eq.~(\ref{eq:mass matrix}).
{Here the indices $p, q$ denote the physical states, $p, q = 1,...,10$ (for a discussion of the elements $O_{pq}$ see for instance \cite{Ghosh:2010zi}}),
and $r$ is a family index.
Thus the entries with the indices $6$ and $7$ correspond in our case to the mixing between neutrinos and neutral higgsinos, $\widetilde H^0_u$ and $\widetilde H^0_ d$, whereas the entries with the index $r$ correspond to the mixing between neutrinos themselves.

{In the following we will assume for simplicity that our sterile neutrino $\nu_s=\nu_{2R}$ mixes only with the active neutrino $\nu_2$.
Then, taking also into account that its mixing with higgsinos is very small~$\sim 10^{-7}$ as discussed in the previous section, one has $O_{si} \to -\frac{1}{2}~U^V_{si}$~and $O_{jk}\to -\frac{1}{2}$~if~$j=k$~and~0 otherwise.} As a result, one obtains
\begin{equation}
    \Gamma(\nu_s \to \nu_i \nu \bar{\nu})=\frac{G_F^2 m_{\nu_S}^5}{32 \pi^3}\ \left(U^V_{si}\right)^2,
    	\label{eq:gamma}
\end{equation}
{where $U^V_{si}$ corresponds to the mixing between sterile and active neutrinos, i.e. $i=2$ in our case.
{The relevance of this result lies in the order of magnitude of the lifetime, long enough for the sterile neutrino to be a good DM candidate.}
Using the typical values 
for $U^V_{si}$ shown in Figure \ref{fig:mixing} of Section~\ref{sec:results}, 
it is straightforward to check that
one obtains a lifetime in the range 
$\tau\sim 10^{24}$~s $- 10^{27}$~s,
which is longer than the age of the
Universe ($\sim10^{17} $~s). Therefore, for this range of masses the sterile neutrino is compatible with a DM particle.

%%%%%%%%%%%%%%%%
% Relic density
%%%%%%%%%%%%%%%%

As suggested in the literature, sterile neutrino DM can be produced via a small mixing with active neutrinos
through the mechanism known as non-resonant
production~\cite{Dodelson:1993je}.
However, this mechanism seems to be ruled out 
(see, e.g. \cite{Boyarsky:2018tvu}) by bounds on active-sterile neutrino mixing and mass bound from structure formation.
%of \cite{Dodelson:1993je}.}
%This
On the other hand, the sterile neutrino DM production can be enhanced by the presence of primordial lepton asymmetry in a scenario known as resonant production~\cite{Shi:1998km}.
The mixing angle determines both the relic density and decay rate, and as a consequence there is a region in the mixing angle-mass parameter space where the sterile neutrino might make up the whole DM. 
In particular, Big Bang Nucleosynthesis (BBN) constraints on the lepton asymetry~\cite{Serpico:2005bc} can be used to set lower limits on the mixing angle, below which DM would be under-produced.
These regions were discussed in detail in 
{Refs.~\cite{Cherry:2017dwu,Perez:2016tcq, Ng:2019gch, Roach:2022lgo}}, including the limits from structure formation and astrophysical X-ray observations (see the discussion in the next sections and Fig.~\ref{fig:mixing}), and we will use their results to constrain our model. To translate them to the $\mn$ cases, one has to take into account that in those works the mixing angle $(U_{si}^V)^2$ is denoted as $\sin^2 \theta$. 

More precise analyses regarding the production 
mechanism\footnote{In addition to the non-resonant and resonant oscillation production models, another interesting alternative is the production via a decaying particle (for a review see e.g. Ref.~\cite{Abazajian:2017tcc} and references therein).} as well as a explicit description of a multiple DM scenario including the sterile neutrino, are beyond the scope of this work, where the latter particle is explored for the first time in the framework of the $\mn$.
%Let us mention nevertheless that the analysis of multiple  DM candidates has already started in this framework~\cite{Gomez-Vargas:2019mqk, Gomez-Vargas:2019vci}.}

%%%%%%%%%%%%%%
% Detection  %
%%%%%%%%%%%%%%
\section{Detection and Exclusion Limits}
\label{sec:xray}
An important constraint on the sterile neutrino comes from its radiative decay $\nu_s \to \nu \gamma$. Although this decay channel is two orders of magitude smaller than the channel into three neutrinos discussed in the previous section, its relevance is due to the produced photons with an energy $m_{\nu_s}/2$ that can be observed in X-ray signals. 
%\st{Also, as it is well known, if the sterile neutrino has $m_{\nu_S}=7$~keV, one could explain the claimed $3.5$~keV X-ray line detection}~\cite{Boyarsky:2014jta,Bulbul:2014sua}.

In the $\mu \nu$SSM, there are several diagrams that contribute to the radiative decay of the sterile neutrino, but only those corresponding to SM processes are relevant. These are shown in Fig.~\ref{fig:wel}, {where mainly the sterile-active neutrino mixing in the incoming line gives rise to the decay of the sterile neutrino via the loop involving a $W$ and two charged fermions.} 
{Note that because of RPV charged leptons have a small mixing with MSSM charginos giving rise to 
five charged fermions $e_{l}$, $l=1, ..,5$.}

\begin{figure}[t!]
    \centering
    \includegraphics[scale=0.5]{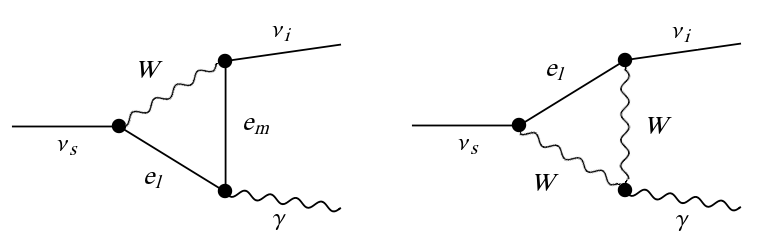}
    \caption{Radiative decay of a sterile neutrino to an active neutrino and a photon via one (two) $W^{\pm}$ boson and two (one) charged fermions $e_{l,m}$ in the loop, with $l,m=1,...,5$.}
    \label{fig:wel}
\end{figure}

\begin{figure}[t!]
    \centering
    \includegraphics[scale=0.5]{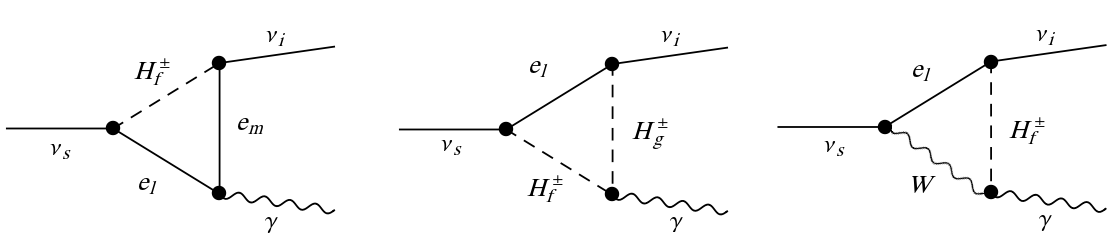}
    \caption{The same as in Fig.~\ref{fig:wel}, but exchanging one or two $W^{\pm}$ bosons in the loop by charged scalars $H^{\pm}_{f,g}$, with $f,g=1,...8$.}
    \label{fig:whpmloop}
\end{figure}

\begin{figure}[t!]
    \centering
    \includegraphics[scale=0.5]{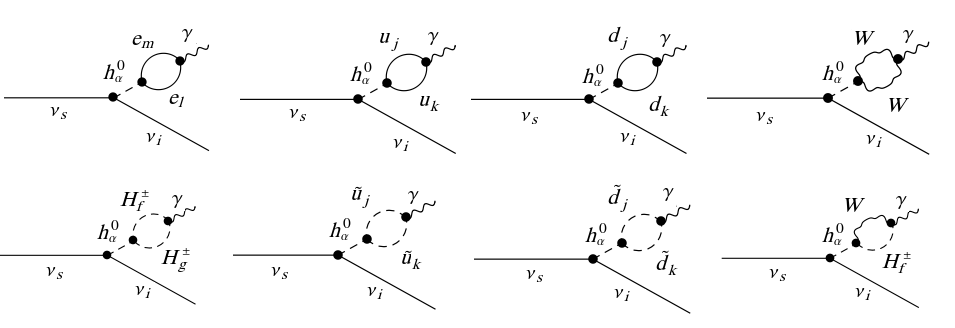}
    \caption{
    Radiative decay of a sterile neutrino to an active neutrino and a photon via neutral scalars.}

    \label{fig:neutral higgs}
\end{figure}

Other diagrams present in the $\mu \nu$SSM are similar to those of Fig.~\ref{fig:wel}, but exchanging in the loop one or two $W^{\pm}$ bosons by charged scalars, as shown in Fig.~\ref{fig:whpmloop}. Again, because of RPV, charged Higgses have a small mixing with right and left charged sleptons giving rise to eight charged scalars
$H^{\pm}_f$, $f=1, ..,8$.
However, all these diagrams are suppressed with respect to those in Fig.~\ref{fig:wel} {because of the small values of the lepton Yukawa coupling in the outgoing vertex and the neutrino Yukawa coupling or the sterile-active neutrino mixing in the incoming vertex.}
{There are also diagrams with the exchange of neutral scalars as shown in 
Fig.~\ref{fig:neutral higgs}, which are also suppressed with respect to those in Fig.~\ref{fig:wel}.  
Although there are more diagrams contributing to this radiative decay, they are even more suppressed than those in Figs.~\ref{fig:whpmloop} and~\ref{fig:neutral higgs} due to the presence of two mixings elements. These diagrams are discussed in Appendix~\ref{appendix}}, together with CP-violating contributions.

To perform the calculation corresponding to the contributions of Fig.~\ref{fig:wel}, we have followed the analysis of Ref.~\cite{Haber:1988px}. There, the decay width of a neutralino to a photon plus a neutrino in the MSSM is computed. {In the case of the $\mn$, it is easy to realize from Fig.~\ref{fig:wel} that the most important contributions have all indices equal, $i=l=m=2$, given that we are working with a sterile neutrino $\nu_s=\nu_{2R}$ that only mixes with $\nu_2$.
Taking it into account, 
we obtain the following result (where the sum over $l$ is implicit):}
\begin{equation}
      \Gamma(\nu_s\to \nu_i \gamma)  = \frac{64 \;   G^2_{\text{F}} \alpha m^3_{\nu_S}}{256 \; \pi^4}
      %\left[\frac{m_{\nu_S}^2e^2g^4}{64\pi^4}
      \left[\frac{3m_{\nu_S}}{4}\left(O_{L,sl}O_{L,il}-O_{R,sl}O_{R,il}\right)- 2m_{e_j}\left(O_{R,sl}O_{L,il}-O_{L,sl}O_{R,il}\right)
      \right]^2
      %\right]
  \label{eq:nugamma}
\end{equation}
where  
\vspace{-0.2cm}
\begin{flalign}
\nonumber O_{R,pl}&=U^V_{p5} U^{e}_{R,l4} +\frac{1}{\sqrt{2}}U^V_{p6}U^{e}_{R,l5}
+\frac{1}{\sqrt{2}}\sum_{r=1}^3U^V_{pr}U^{e}_{R,lr}, \\
O_{L,pl}&=U^{V}_{p5}U^{e}_{L,l4}-\frac{1}{\sqrt{2}}U^{V}_{p7}U^{e}_{L,l5}, 
\label{eq:nugamma2}
\end{flalign}
with $U^e_R$ and $U^e_L$ the matrices which diagonalize the mass matrix for the charged fermions 
{(the elements $O_{R,pl}$ and $O_{L,pl}$ were discussed in Ref.~\cite{Ghosh:2010zi})}.
For these matrices, the entries with the indices 4 and 5 correspond to the small mixing between  charged leptons and charged wino and higgsino, whereas the entries with the index $r$
correspond to the mixing between charged leptons themselves.
Concerning the matrix $U^V$, the entries with the index 5 correspond to the mixing between neutrinos and neutral wino~$\lsim 10^{-7}$, whereas the entries with the index $r$ correspond to the mixing between neutrinos themselves.
Thus we can neglect the products $U^V U^e$ in Eq.~(\ref{eq:nugamma2}),
obtaining {$O_{L,il}=O_{L,sl}\rightarrow 0$} and 
 $O_{R,sl}O_{R,il} \to \frac{1}{2}~U^V_{si}$. 
Therefore, the corresponding decay width is
\begin{equation}
    \Gamma(\nu_s\to \nu_i \gamma) = \frac{9G_F^2\alpha m_{\nu_S}^5 }{256 \pi^4}\ \left(U^V_{si}\right)^2,
\label{eq:decaynugamma}
\end{equation}
where as for the case of the three-body decay in the previous section, $U^V_{si}$ corresponds to the mixing between sterile and active neutrinos with $i=2$.
This result agrees with the one of Ref.~\cite{Boyarsky:2009ix}, where this process is calculated in an extension of the SM with RH neutrinos. 

Several constraints on the active-sterile neutrino mixing coming from this radiative decay, can be found in the literature. In particular, this is the case of
X-ray signals due to a DM flux that contributes to the X-ray background (XRB) from galaxy clusters, dSPh galaxies, and Milky Way halo, among others~\cite{Boyarsky:2005us,Abazajian:2001vt, Bulbul:2014sua, Horiuchi:2013noa, Boyarsky:2007ay, Boyarsky:2007ge, Watson:2011dw, Abazajian:2006jc, Boyarsky:2006zi, Boyarsky:2006ag, Loewenstein:2008yi,Loewenstein:2009cm, Malyshev:2014xqa, Boyarsky:2006kc, Ng:2015gfa, Foster:2021ngm}. 
As mentioned in the previous section, these constraints are summarized in {Refs.~\cite{Cherry:2017dwu, Perez:2016tcq, Ng:2019gch, Roach:2022lgo}}, and we will take them into account in our numerical analysis of the results below.}

%%%%%%%%%%%%%%%%%%%%%%%%%%%%%%%%%%%%%%%%%%%%%%%%
%%%%%%%             Results            %%%%%%%%%
%%%%%%%%%%%%%%%%%%%%%%%%%%%%%%%%%%%%%%%%%%%%%%%%     
\section{Results} 
\label{sec:results}

We present here the numerical results obtained using the formulas discussed in previous sections. To find first 
a sterile neutrino state fulfilling neutrino physics~\cite{Capozzi:2017ipn,deSalas:2017kay,deSalas:2018bym,Esteban:2018azc},
we used SARAH~\cite{Staub:2013tta} to generate a 
{\tt SPheno}~\cite{Porod:2003um,Porod:2011nf} version for the model. 
 {We also checked that our results} 
 %\st{benchmark points} 
 satisfy the constraints coming from the Higgs sector.
 In particular, we implemented LEP, LHC and Tevatron Higgs constraints using \texttt{HiggsBounds} {{v}}5.9.1~{\cite{Bechtle:2008jh,Bechtle:2011sb,Bechtle:2013gu,Bechtle:2013wla,Bechtle:2015pma,Bechtle:2020pkv}}, and we used  \texttt{Higgssignals} {{v}}2.6.1~\cite{Bechtle:2013xfa,Stal:2013hwa,Bechtle:2014ewa,Bechtle:2020uwn} in order to compute a $\chi^2$ measure to determine the compatibility of our model with the measured signal strength and mass. We required that the $p$-value reported by \texttt{Higgssignals} be larger than 
 $5\%$.

 \begin{figure}[t!]
     \centering
%\hspace*{-1.5cm}
%        \includegraphics[scale=0.58]{figures/Figure_final_linear.png}
     \includegraphics{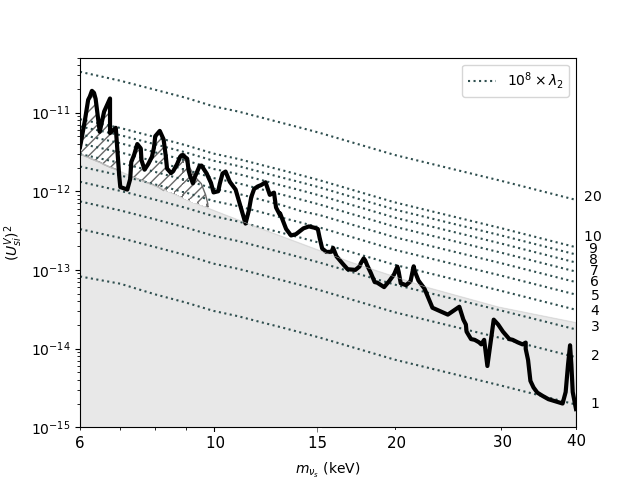}
     \caption{Sterile-active neutrino mixing $(U^V_{si})^2$ versus sterile neutrino mass $m_{\nu_s}$. Dotted
     lines show the allowed values of $(U^V_{si})^2$
     %the active-sterile mixing 
     for several fix values of $\lambda_2$ in the range $(1-20)\times 10^{-8}$. The region above the black line is forbidden by X-ray constraints,
%     the experimental constraints available, 
     as summarized in Ref.~\cite{Roach:2022lgo}. 
     %Above this line the region is experimentally forbidden. 
     The gray hatched region represents structure-formation constraints from the observed number of Milky Way dwarf satellites \cite{Cherry:2017dwu}.
     %, again as used in \cite{Roach:2022lgo}. 
     In the gray region BBN constraints imply that DM would be under produced. 
     }
     \label{fig:mixing}
 \end{figure}

%\begin{table}[t!]
%\begin{center}
%\begin{tabular}{|c| c| c |c |c |c |}
%\hline 
%$v_{1L}$ & $v_{2L}$ &  $v_{3L} $ & $v_{1R}$ & $v_{2R}$  & $v_{3R}$  \\
%\hline
%$1.00\times10^{-4}$  & $4.45\times10^{-4}$ & $3.80\times10^{-4}$ & $2.40\times10^3$ & $1.50\times10^3$ & $1.40\times10^3$ \\
%\hline
%\end{tabular}
%\end{center}
%\caption{Values of the input left and right sneutrino VEVs (in GeV) that are fixed in the computation.}
%\label{table:vevs}
%\end{table}

%\begin{table}[t!]
%\begin{center}
%\begin{tabular}{|c| c| c |c |c| c|}
%\hline 
%$\lambda_1$ &  $\lambda_3$  & $\kappa_{1}$ & $\kappa_{3}$ & $M_1$ & $M_2$\\
%\hline
%$4.0\times10^{-4}$ &  $1.5\times10^{-1}$ & $1.0\times10^{-1}$ & $4.5\times10^{-1}$ & 360 & 1600\\
%\hline
%\end{tabular}
%\end{center}
%\caption{Values of the input couplings $\lambda_1, \lambda_3, \kappa_1, \kappa_3$, and input bino and wino masses $M_1, M_2$ (in GeV) that are fixed in the computation.}
%\label{table:lambda}
%\end{table}

%%% reescribir

%% MOVIDO

Concerning the sterile neutrino DM,
in Fig~\ref{fig:mixing} we show the current limits on the parameter space. The region above the black line is forbidden by X-ray constraints, and the gray hatched region represents structure-formation constraints. Imposing the value of the relic density~\cite{Cherry:2017dwu} (above the gray region), we can see that
the allowed sterile neutrino masses are in the range
$m_{\nu_s}\sim {10} - 15$ keV, corresponding to a sterile-active neutrino mixing
$(U^V_{si})^2\sim 2\times 10^{-12} - 2\times 10^{-13}$. 
Using the parameter space of the $\mn$, we are able to reproduce these ranges of values.

\begin{table}[t!]
\begin{center}
\begin{tabular}{|c| c| c |c |c |c |}
\hline 
$v_{1L}$ & $v_{2L}$ &  $v_{3L} $ & $v_{1R}$ & $v_{2R}$  & $v_{3R}$  \\
\hline
$1.00\times10^{-4}$  & $4.45\times10^{-4}$ & $3.80\times10^{-4}$ & $2.40\times10^3$ & $1.50\times10^3$ & $1.40\times10^3$ \\
\hline
\end{tabular}
\end{center}
\caption{Values of the input left and right sneutrino VEVs (in GeV) that are fixed in the computation.}
\label{table:vevs}
\end{table}

\begin{table}[t!]
\begin{center}
\begin{tabular}{|c| c| c |c |c| c|}
\hline 
$\lambda_1$ &  $\lambda_3$  & $\kappa_{1}$ & $\kappa_{3}$ & $M_1$ & $M_2$\\
\hline
$4.0\times10^{-4}$ &  $1.5\times10^{-1}$ & $1.0\times10^{-1}$ & $4.5\times10^{-1}$ & 360 & 1600\\
\hline
\end{tabular}
\end{center}
\caption{Values of the input couplings $\lambda_1, \lambda_3, \kappa_1, \kappa_3$, and input bino and wino masses $M_1, M_2$ (in GeV) that are fixed in the computation.}
\label{table:lambda}
\end{table}

As discussed in Section~\ref{sec:munu}, the relevant parameters of the model in the neutrino sector are those given in Eq.~(\ref{eq:independent params}). Three
of them, namely 
$Y^{\nu}_{22}$,
$\kappa_{2}$ and $\lambda_2$ are crucial to obtain a light sterile neutrino.
For the rest of the parameters we have chosen values that are characteristic of the model. This is the case of the values shown in 
Tables~\ref{table:vevs} and~\ref{table:lambda}.
For simplicity, in the latter table we chose $\lambda_3$ to dominate the contribution to the $\mu$-term as well as the mixing with Higgsinos, but similar values for $\lambda_1\sim 10^{-1}$ are also possible without modifying significantly our results.
%the values of the rest of the parameters, keeping our results general.
The same comment applies to $\tan\beta$, although we chose the common value
$\tan\beta = 10$, other values can be used keeping our results general.
%would not modify our results.
Concerning the neutrino Yukawas,
%We also fixed
%$\tan\beta = 10$  which is a common value. A different election for
%$\tan\beta$ \R{(or $\lambda_1 \sim \lambda_3 \sim 10^{-1}$)} %\M{(or the mentioned absent of hierarchy between $\lambda_1$ and  $\lambda_3$)  } 
%would not modify significantly the values for the rest of the parameters, keeping our results general.
we have to work with off-diagonal entries in the 
%neutrino 
Yukawa matrix to obtain the correct mixing angles. The values of $Y^{\nu}_{22}$ and the other entries of the matrix used in our computation are:
\begin{equation}
    Y^{\nu}= \begin{pmatrix}
8.0\times10^{-8} & 0 & 1.7\times10^{-8} \\
{-2.8\times10^{-7}} & 4.0\times10^{-13} & {-8.6\times10^{-8}} \\
0 & 0 & 5.5\times10^{-8} 
\end{pmatrix},
\label{eq:yukawa}
\end{equation}
where as chosen in previous sections the sterile neutrino $\nu_s=\nu_{2R}$ mixes only with the active neutrino $\nu_2$.
Different values chosen for $\lambda_2$ are shown in Fig~\ref{fig:mixing}
together with the corresponding 
values of the mixing $(U^V_{si})^2$ (taking $i=2$ in our analysis), and
sterile neutrino mass $m_{\nu_s}$. The latter mass fixes the value of $\kappa_{2}$ from~(\ref{eq:sterile mass}).
As we can see,
for a given value of $\lambda_2$ the mixing becomes smaller with increasing mass. This is a reasonable result since for larger masses the sterile and active neutrino mass difference increases, giving rise to a smaller mixing element.
Finally, concerning the scalar and pseudoscalar sectors, as discussed in Section~\ref{sterile}     
$A_{\lambda}$ and $A_{\kappa}$ are relevant parameters for them, and we used in all cases the representative values
 $A^{\lambda}_1=250~$GeV, $A^{\lambda}_2=5~$TeV, $A^{\lambda}_3=666~$GeV, $A^{\kappa}_1=-2.2~$GeV, $A^{\kappa}_2=-0.23~$GeV and $A^{\kappa}_3=-9.4~ $GeV.

We conclude from the results of Fig~\ref{fig:mixing}, and taking into account reasonable variations of the parameters chosen, that 
%an order of magnitude of
$\lambda_2$ has to be in the range 
$\lambda_2\sim  10^{-7} - 10^{-8}$
%is necessary
in order 
to reproduce the value of the relic density.

On the other hand, if one allows the sterile neutrino to produce a lower value of the relic density below the observational bound, all the mass range of 
Fig~\ref{fig:mixing} will be available for appropriate values of $\lambda_2$ in the $\mn$.
We checked that this is also true for sterile neutrino masses larger than 40 keV. In particular, 
for cases with
$m_{\nu_s}\gsim 50$ keV we considered in our analysis the additional constraints discussed in Ref.~\cite{Laha:2020ivk}, where an study of the INTEGRAL data was carried out.
Of course, for all these cases an additional contribution/s from another/other DM candidate/s would be necessary in order to obtain the relic density in the observational range.
Besides, it is
also possible to find points 
in the $100$ keV$-$1 MeV region, still in agreement with a lifetime longer than the age of the Universe.
For instance, working with similar parameters as in %all the examples in the keV range given in 
Tables~\ref{table:vevs} and~\ref{table:lambda},
and Eq.~(\ref{eq:yukawa}), used for the keV mass region, but with $\kappa_2=1.42 \times 10^{-7}$ and $\lambda_2=7 \times 10^{-11}$, we obtain a sterile neutrino with $m_{\nu_s}=300$~keV, $(U^V_{si})^2 \sim  10^{-21}$, and according to Eq.~(\ref{eq:gamma}) a lifetime $\tau \sim 10^{27}$~s.
{In~addition~to those constraints in
Ref.~\cite{Laha:2020ivk} mentioned above, more constraints on the active-sterile neutrino mixing can be found in the literature for this mass range.}
They arise from nuclear beta decay and pion decay, among others, corresponding to the mass range $\sim 300$ keV$-$1000 MeV, as summarized in Ref.~\cite{Bryman:2019bjg}.
Although producing upper
bounds on the mixing angle,
%they
{all constraints available in this region}
are easily fulfill by our results.

To conclude our discussion, in order to cover a wider mass range for the sterile neutrino, we searched for points {above} the MeV scale, with $m_{\nu_s}=10$ MeV and 100 MeV. In these cases the sterile neutrino lifetimes are smaller than the age of the Universe and it cannot be a DM candidate.
For them we used the Yukawa matrix of Eq.~(\ref{eq:yukawa}) and the values of the parameters of Tables \ref{table:vevs} and \ref{table:lambda}.
Concerning the experimental constraints, we applied again those 
of Ref.~\cite{Bryman:2019bjg}, with the result that the upper bounds on the mixing angle are fulfilled by the typical values obtained in our computation. In particular, for the analyzed range of masses, using $\lambda_2 \sim 10^{-6} - 10^{-4}$ and $\kappa_2 \sim 10^{-6} - 10^{-5}$ 
we obtained $(U^V_{si})^2 \sim 10^{-18} - 10^{-12}$, which are values far from the experimental available constraints.
%These values correspond to $\lambda_2 \sim 10^{-6} - 10^{-4}$ and $\kappa_2 \sim 10^{-6} - 10^{-5}$ }

%%%%%%%%%%%%%%%%%%%%%%%%%%%%%%%%%%%%%%%%%%%%%%%%
%%%%          Conclusions                   %%%%
%%%%%%%%%%%%%%%%%%%%%%%%%%%%%%%%%%%%%%%%%%%%%%%%
\section{Conclusions}
\label{sec:conclusions}
Neutrino masses and mixing angles in the $\mn$ arise naturally from a generalized EW-scale seesaw mechanism {involving the neutralinos in addition to RH and LH neutrinos. This is obtained without the need to introduce additional scales beyond the soft SUSY-breaking scale.
The presence of RH neutrino superfields in the model is crucial to solve not only this $\nu$-problem, but also the $\mu$-problem of SUSY models.
In this work, we carried out the first analysis in the $\mn$ of using RH neutrinos as sterile neutrinos. In particular, two RH neutrinos were used to generate masses at tree level for the active neutrinos through the
generalized seesaw, whereas 
the third one plays the role of a sterile neutrino by choosing appropriately the parameters of the model.

We analysed numerically this parameter space, reproducing the current data on neutrino physics for sterile neutrino masses in the range $6~\text{keV}< m_{\nu_s} < 40~\text{keV}$.
%as shown in Fig.~\ref{fig:mixing}. 
%Tables~\ref{table:kappa1} and~\ref{table:kappa2}. 
{All the points obtained fulfil also collider constraints coming from the Higgs sector.}
Besides, for this range of masses the relevant decay mode $\nu_s \to \nu \nu \nu$ gives rise to a lifetime for the sterile neutrino which lies between about  
{$\tau\sim 10^{24} - 10^{27}$~s}, longer than the age of the Universe ($\sim10^{17} $~s). Therefore, it is compatible with a DM particle. 
%\st{Because of the presence of the radiative decay channel $\nu_s \to  \nu \gamma$, we checked that these points also fulfill the limits from astrophysical X-ray observations. }
%\st{only those points} \st{in bold in Table}~\ref{table:kappa1} 
As shown in Fig. \ref{fig:mixing}, only specific values of the relevant parameter $\lambda_2$, which couples the Higgses with the sterile neutrino, are
in agreement with the bounds imposed by the relic density and  X-ray observations
(because of the radiative decay channel $\nu_s \to  \nu \gamma$) 
corresponding to the 
range of masses $m_{\nu_s}\sim {10} - 15$ keV.

On the other hand,
the whole mass range of
Fig. \ref{fig:mixing} 
for appropriate values of $\lambda_2$, and also larger masses, can be in agreement with X-ray observations if one allows a value of the relic density below the observational bound. In this case, the presence of other DM candidates is mandatory.
We also checked that it is possible to find points in the 100 keV$-$1 MeV region, still with the lifetime of the sterile neutrino longer than the age of the Universe.
In this region the current experimental constraints on the active-sterile mixing angle are also fulfilled.
Finally, we searched for viable parameters giving sterile neutrino masses above the MeV scale,
where this particle
cannot be a DM candidate, finding compatibility with the experimental constraints.

%%%%%%%%%%%%%%%%%%%%%%%%%%%%%%%%%%%%%%%%%%%%%%%%

%%%%%%%%%%%%%%%%%%%%%%%%%%%%%%%%%%%%%%%%%%%%%%%%%%%%%%%%%%%%%%%%%%%%%%%%%%

%%%%%%%%%%%%%%%%%%%%%%%%%%%%%%%%%%%%%%%%%%%%%%%%%%%%%%%%%%%%%%%%%%%%%%%%%%
%%%%%%%%%%%%         Acknowledgments              %%%%%%%%%%%%%%%%%%%%%%%%
%%%%%%%%%%%%%%%%%%%%%%%%%%%%%%%%%%%%%%%%%%%%%%%%%%%%%%%%%%%%%%%%%%%%%%%%%%
%\pagebreak
%\clearpage
\begin{acknowledgments}

{CM thanks Werner Porod for useful comments.}
{PK thanks Essodjolo Kpatcha for help with code implementation}. The work of PK and DL was supported by the Argentinian CONICET, and they also acknowledge the support through PIP11220170100154CO and {PICT~2020-02181}.  
The research of CM was partially supported by the Spanish Research~Agency~(AEI)
through the grants IFT Centro de Excelencia Severo Ochoa No CEX2020-001007-S,
PGC2018-095161-B-I00, and PID2021-125331NB-I00. They are funded by MCIN/AEI/10.13039/501100011033, and the
second grant also by ERDF "A way of making Europe".

\end{acknowledgments}
%%%%%%%%%%%%%%%%%%%%%%%%%%%%%%%%%%%%%%%%%%%%%%%%%%%%%%%%%%%%%%%%%%%%%%%%%%

%%%%%%%%%%%%%%%%%%%%%%%%%%%%%%%%%%%%%%%%%%%%%%%%%%%%%%%%%%%%%%%%%%%%%%%%%%
%%%%%%%%%%%%         Appendix                     %%%%%%%%%%%%%%%%%%%%%%%%
%%%%%%%%%%%%%%%%%%%%%%%%%%%%%%%%%%%%%%%%%%%%%%%%%%%%%%%%%%%%%%%%%%%%%%%%%%
%\clearpage
\pagebreak
\appendix
\section{Appendix} \label{appendix}

RPV in the $\mu \nu$SSM is related to neutrino physics, and as a consequence the one loop diagrams of the process $\nu_s\to \nu \gamma$ are suppressed by small Yukawas and/or small admixtures. 
The dominant ones were discussed in Fig.~\ref{fig:wel}, and
those in Figs.~\ref{fig:whpmloop} and~\ref{fig:neutral higgs} are more suppressed.
For completeness we discuss here those that follow in relevance.
{They are shown in 
Fig.~\ref{fig:loop quark-squark}.} 

As we can deduce from the figure,
there are two mixing elements present. One concerns the incoming sterile neutrino mixing with a higgsino, which in turn decays to $q \tilde{q}$. This mixing is~$\sim10^{-7}$. The second one concerns an outgoing higgsino mixing with an active neutrino, which can be even {smaller down to~$\sim 10^{-11}$.}

\bigskip

\noindent
Finally,
although in this work we have assumed CP conservation for simplicity, it is worth noting that all CP-violating contributions are 
suppressed {similarly to those contributions in Fig.~\ref{fig:neutral higgs}}. They are shown in 
Fig.~\ref{fig:pseudoscalar loop}.

\begin{figure}[b!]
    \includegraphics[scale=0.55]{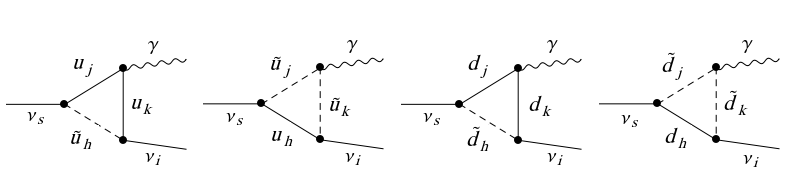}
    \caption{Radiative decay of a sterile neutrino to an active neutrino and a photon via quark/squark loops, with $j,k,h=1,2,3$. }
    \label{fig:loop quark-squark}
\end{figure}

\begin{figure}[b!]
    \centering
    \includegraphics[scale=0.5]{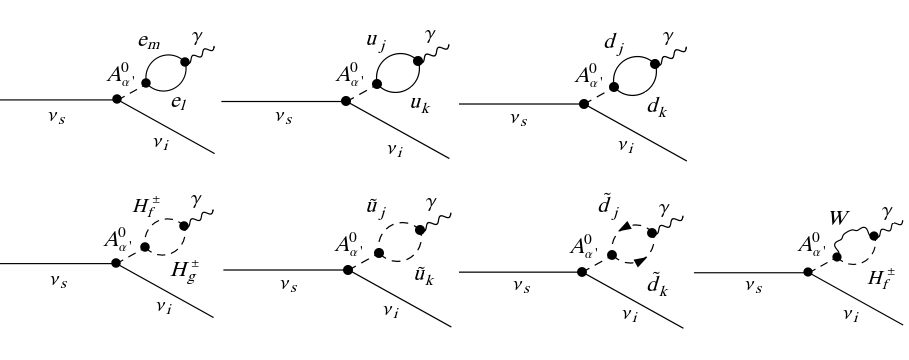}
    \caption{Radiative decay of a sterile neutrino to an active neutrino and a photon via pseudoscalars.} 
    %\R{Similares dudas que en figura 4. Se sale la cuarta figura de abajo de los margenes.}
    \label{fig:pseudoscalar loop}
\end{figure}

%%%%%%%%%%%%%%%%%%%%%%%%%%%%%%%%%%%%%%%%%%%%%%%%%%%%%%%%%%%%%%

%%%%%%%%%%%%%%%%%%%%%%%%%%%%%%%%%%%%%%%%%%%%%%%%%%%%%%%%%%%%%

%%%%%%%%%%%%%%%%%%%%%%%%%%%%%%%%%%%%%%%%%%%%%%%%%%%%%%%%%%%%%%%

%%%%%%%%%%%%%%%%%%%%%%%%%%%%%%%%%%%%%%%%%%%%%%%%%%%%%%%%%%%%%%%%%%%%%%%%%%

%%%%%%%%%%%%%%%%%%%%%%%%%%%%%%%%%%%%%%%%%%%%%%%%%%%%%%%%%%%%%%%%%%%%%%%%%%
%%%%%%%%%%%%         Bibliography                 %%%%%%%%%%%%%%%%%%%%%%%%
%%%%%%%%%%%%%%%%%%%%%%%%%%%%%%%%%%%%%%%%%%%%%%%%%%%%%%%%%%%%%%%%%%%%%%%%%%
\clearpage
\bibliographystyle{utphys}
%\bibliography{munussmbib-completo_v6}
\bibliography{munu-steril-DM.bbl}

%%%%%%%%%%%%%%%%%%%%%%%%%%%%%%%%%%%%%%%%%%%%%%%%%%%%%%%%%%%%%%%%%%%%%%%%%%

\end{document}